\documentclass{article}

 \usepackage[final,nonatbib]{neurips_2025_ml4ps}

\usepackage[utf8]{inputenc} 
\usepackage[T1]{fontenc}    
\usepackage{hyperref}       
\usepackage{url}            
\usepackage{booktabs}       
\usepackage{amsfonts}       
\usepackage{nicefrac}       
\usepackage{microtype}      
\usepackage{xcolor}         
 \usepackage{tikz} 
 \usetikzlibrary{shapes.geometric, arrows, positioning, fit, calc}
 \usepackage[backend=biber,sorting=none,style=numeric]{biblatex}
 \usepackage{graphicx}
\usepackage{float} 
\usepackage{booktabs}
\usepackage{multirow}
\title{MITRA: An AI Assistant for Knowledge Retrieval in Physics Collaborations}

\author{%
  Abhishikth Mallampalli\\
  University of Wisconsin-Madison\\
  \texttt{amallampalli@wisc.edu} \\
   \And
   Sridhara Dasu \\
   University of Wisconsin-Madison\\
   \texttt{dasu@hep.wisc.edu} \\
}

\begin{document}

\maketitle

\begin{abstract}
  Large-scale scientific collaborations, such as the Compact Muon Solenoid (CMS) at CERN, produce a vast and ever-growing corpus of internal documentation. Navigating this complex information landscape presents a significant challenge for both new and experienced researchers, hindering knowledge sharing and slowing down the pace of scientific discovery. To address this, we present a prototype of MITRA, a Retrieval-Augmented Generation (RAG) based system, designed to answer specific, context-aware questions about physics analyses. MITRA employs a novel, automated pipeline using Selenium for document retrieval from internal databases and Optical Character Recognition (OCR) with layout parsing for high-fidelity text extraction. Crucially, MITRA's entire framework, from the embedding model to the Large Language Model (LLM), is hosted on-premise, ensuring that sensitive collaboration data remains private. We introduce a two-tiered vector database architecture that first identifies the relevant analysis from abstracts before focusing on the full documentation, resolving potential ambiguities between different analyses. We demonstrate the prototype's superior retrieval performance against a standard keyword-based baseline on realistic queries and discuss future work towards developing a comprehensive research agent for large experimental collaborations.

\end{abstract}

\section{Introduction}

Experimental physics collaborations today are larger and more productive than ever, leading to an incredible and accelerating pace of scientific discovery. This success, however, brings a new challenge: managing an ever-growing mountain of internal knowledge. Collaborations like CMS consist of thousands of members who generate terabytes of data and a corresponding volume of documentation, including analysis notes, internal wikis, and procedural guidelines. For a new PhD student joining an analysis group or an expert seeking to quickly understand the nuances of a specific measurement, finding precise information can be a time-consuming and often frustrating endeavor. Traditional keyword-based search tools often fail to capture the semantic context of a query and and are heavily dependent on the exact phrasing match between query and the language in the text. A significant amount of expert effort is invested in producing these detailed documents, yet their contents are often underutilized due to inefficient information retrieval.

To address this challenge, this work introduces a prototype of MITRA (a Sanskrit word meaning 'friend'), a conversational AI system designed to serve as an expert assistant for members of a large physics collaboration. We leverage the power of Retrieval-Augmented Generation (RAG)~\cite{lewis2020retrieval} to build a system that can provide accurate, cited answers to natural language questions about specific physics analyses. Concurrent with our work, other collaborations are also exploring similar solutions. For instance, the ATLAS collaboration is developing a system that relies on external, API-based services (e.g., OpenAI's GPT-4o mini~\cite{openai_pricing_2025}), citing cost-versus-quality considerations~\cite{DalSanto:2935252}. While this API model is highly cost-effective for individual queries, the cumulative operational expense for a collaboration with thousands of members over several years can become substantial. Our work presents a complementary, privacy-first alternative that also offers a more sustainable financial model. By demonstrating a fully on-premise framework that leverages existing institutional hardware, we avoid ongoing per-token costs and, most critically, ensure that no proprietary research data ever leaves the local servers. Our primary contributions are:
\begin{itemize}
    \item A modular and automated pipeline for ingesting and processing internal analysis documents, designed to support extensibility and version-aware updates.
    \item A design philosophy centered on a two-tiered database system to maintain context and avoid cross-analysis confusion.
    \item A fully on-premise deployment model that guarantees the privacy and security of proprietary collaboration data.
\end{itemize}

\section{Methodology: An Automated and Private Pipeline}

MITRA is built on a robust pipeline that transforms raw documents from a database into a queryable knowledge base. Since analysis documents are frequently updated, this RAG architecture allows for the knowledge base to be refreshed by simply re-embedding the updated documents --- a process far more efficient than the costly retraining required for monolithic language models.

\subsection{Automated Document Acquisition and Extraction}
To access the analysis documents, which are often stored in databases with web interfaces, we utilize \texttt{Selenium}~\cite{selenium}, a browser automation tool. Scripts are developed to log into the system, navigate to the relevant sections, and download the analysis notes, typically in PDF format.

Once retrieved, the documents are processed. Instead of relying on standard PDF-to-text libraries (e.g., \texttt{PyPDF}~\cite{pypdf}, \texttt{PDFPlumber}~\cite{pdfplumber}), which can struggle with complex layouts, we employ advanced Optical Character Recognition (OCR) engines (e.g., \texttt{Surya}~\cite{paruchuri2025surya} and \texttt{Tesseract}~\cite{smith2007tesseract}). Modern OCR tools offer superior layout parsing capabilities, allowing us to accurately extract text while preserving the distinction between main content, figure captions, page numbers, line numbers and tables. This high-fidelity extraction is critical for creating a high-quality knowledge base.

\subsection{Embedding, Retrieval, and Reranking}
The extracted text is segmented into manageable chunks (e.g., paragraphs or sections). For our prototype, we chose to chunk by paragraph, as this often aligns with the logical separation of ideas in the source documents. To create a searchable knowledge base, we use a Dense Passage Retrieval (DPR) model to encode these document chunks into a 768-dimensional vector space. We specifically use the \texttt{facebook/dpr-question\_encoder-multiset-base} model~\cite{karpukhin2020dense}, which has demonstrated strong performance on public benchmarks, via the \texttt{Transformers} library~\cite{wolf-etal-2020-transformers}. These embeddings are then stored in a lightweight and efficient open-source vector database such as Chroma DB~\cite{chroma}. When a user asks a question, the same DPR model encodes the query. The initial retrieval step performs a similarity search, using cosine similarity~\cite{cosine_sim}, to find the top-k most relevant passages from the database, where k is a tunable parameter. A larger k increases the computational load of the subsequent reranking step but raises the probability of capturing the most relevant passage. These candidates are then passed to a more computationally intensive but accurate reranking step. For this, we use a cross-encoder model (\texttt{cross-encoder/ms-marco-MiniLM-L-6-v2})~\cite{reimers-2019-sentence-bert,Bajaj2016Msmarco,wang2020minilm}, which scores the relevance of each passage to the query with higher accuracy. While computationally expensive, this two-stage process is necessary because cross-encoders, which jointly process query and passage pairs, are too slow for an initial search over the entire database. The final ranked list of passages is then used as context for the LLM. To mitigate the risk of model hallucination, the LLM is explicitly prompted to ground its answers strictly within the retrieved context passages.

\subsection{On-Premise LLM for Privacy}
A core principle of our design is data privacy. All components of MITRA are hosted on local collaboration GPU servers (e.g., NVIDIA Tesla T4 with 15GB of memory). This includes the embedding model and, most importantly, the LLM. We use a 4-bit quantized version of \texttt{Mistral-7B} model~\cite{jiang2023mistral}, which allows the 7.2B parameter model to run efficiently on our hardware while maintaining strong performance. The LLM is served locally using \texttt{Ollama}~\cite{ollama} and integrated into our application pipeline with the \texttt{LangChain} framework~\cite{langchain}. This on-premise approach provides two key benefits. First, for a large collaboration with thousands of members, it avoids the substantial cumulative cost that API-based services would incur over time. Second, and most critically, it ensures that no proprietary analysis details, unpublished results, or internal documentation are ever transmitted outside the collaboration's secure network.

\begin{figure}[H]
    \centering 
    
    \includegraphics[width=0.75\textwidth]{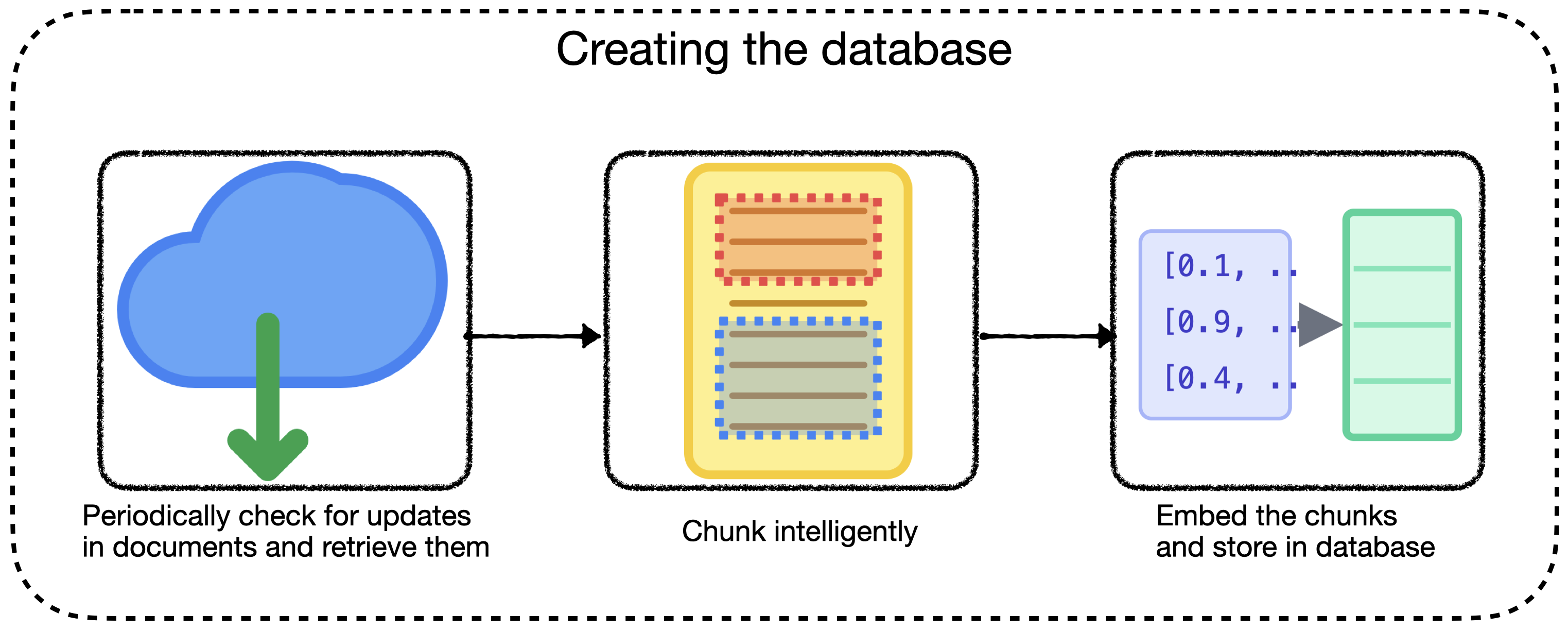}
    
    \includegraphics[width=0.8\textwidth]{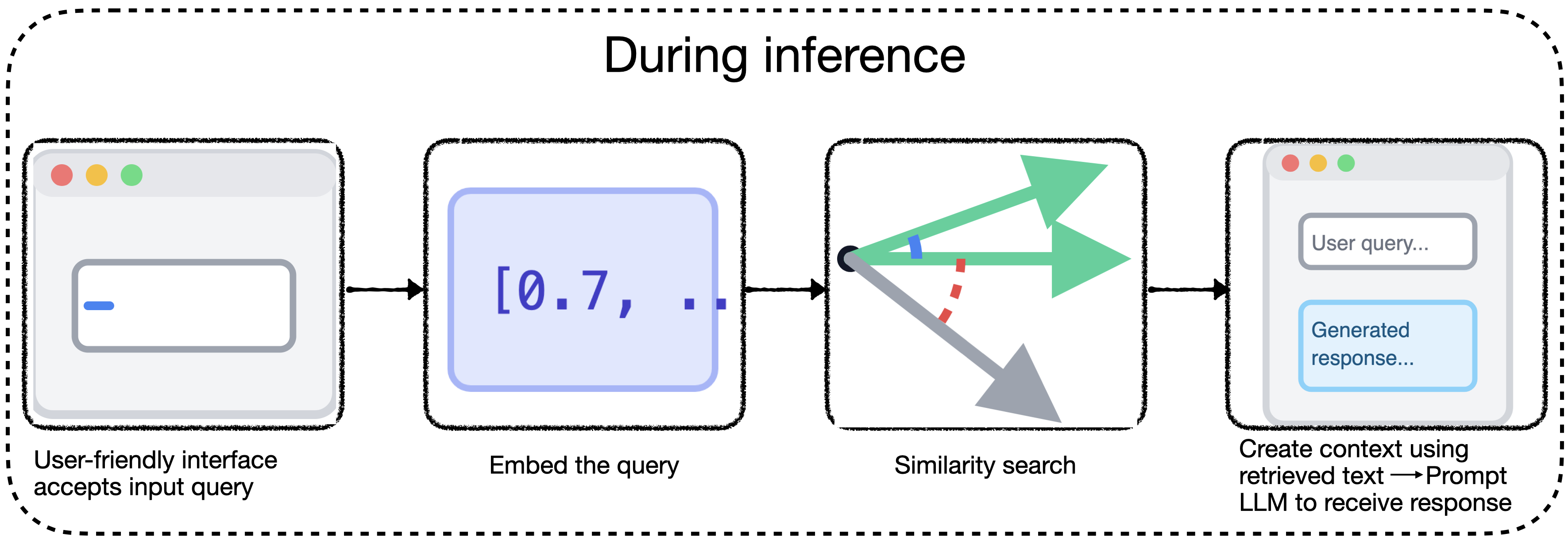}
    
    \caption{Workflow of MITRA, divided into the offline database creation pipeline (top) and the online inference process (bottom).}
    \label{fig:two_images}
\end{figure}

\section{System Design: A Two-Tiered Database Approach}
A key challenge in answering analysis-specific questions is that the ``correct" answer is often context-dependent. For example, the question ``What is the most important background?" will have a different, and potentially conflicting, answer for a Higgs to di-muon analysis versus a search for dark matter.

To resolve this, we implemented a two-tiered database system:
\begin{enumerate}
    \item \textbf{Abstracts Database:} This primary database contains only the abstracts from all available analyses documents. When a user initiates a conversation, their first query is used to perform a similarity search against this database. The goal is to identify the single analysis most relevant to the user's question. The system then prompts the user to confirm this choice. This human-in-the-loop validation step is critical, as it ensures the correct document context is established before proceeding and prevents the system from being locked into an irrelevant analysis.
    \item \textbf{Full-Text Database:} Once the top-ranked analysis is identified from the abstracts, the system ``locks on" to a second, dedicated database containing the full, chunked text of only that specific analysis. All subsequent queries and RAG operations within that conversation session are directed exclusively to this single-analysis database. Every single-analysis database consists of information from multiple documents related to that analysis and this combined context is passed to the generator. This session-specific context allows users to start a new conversation at any time to ask about a different analysis.
\end{enumerate}

This approach effectively isolates the context, ensuring that the system's responses are consistently grounded in the correct document and preventing the model from conflating information from different analyses. The user interacts with the system via a clean, intuitive web interface built with \texttt{Streamlit}~\cite{streamlit}.

\section{Performance Evaluation}
To quantitatively evaluate the effectiveness of our semantic retrieval pipeline, we compare its performance against a standard keyword-based baseline, Okapi BM25~\cite{BM25}. BM25 is a robust ranking function based on term frequency and inverse document frequency that is widely used in search engines. It excels at matching exact keywords but does not inherently understand the semantic meaning or context of the query.

We conducted the evaluation using two sets of queries designed by domain experts to be representative of real-world information needs. ``Set 1" consists of queries using the exact phrasing found within the source documents. ``Set 2" contains more realistic queries, where the user's phrasing uses common synonyms or paraphrases the concepts (e.g., asking for ``transverse momentum requirement" when the text uses ``\(p_T\) cut"). Performance is measured using standard information retrieval metrics: Precision@k (P@k), which is the fraction of relevant text sections in the top-k results, and Recall@k (R@k), the fraction of all relevant text sections found in the top-k results.

The results, summarized in Table~\ref{tab:rag-evaluation-1} and Table~\ref{tab:rag-evaluation-2}, highlight the critical advantage of MITRA. 

The results in Table~\ref{tab:rag-evaluation-1} show that for Set 1 queries with exact keyword matches, both systems perform strongly \footnote{Note that Precision@k naturally decreases as k increases in our setting because most queries have only one relevant answer. If all queries had only one relevant answer then the ideal P@5 would be 0.2. Similarly, Recall@k naturally increases as k increases because more passages are retrieved.}. As expected, the BM25 baseline is highly effective at this task, slightly outperforming the RAG agent on the P@5 and R@5 metrics due to its strength in exact keyword matching. However, for the more realistic Set 2 queries, the performance of BM25 drops significantly. MITRA, in contrast, maintains high performance, achieving a Precision@1 of 0.75 compared to just 0.13 for BM25. This demonstrates that our search strategy correctly understands the semantic equivalence between the user's query and the document's language, successfully retrieving the correct passages even without an exact keyword match. This capability is essential for a practical, user-friendly system where users cannot be expected to know the precise terminology of the source documents.

\begin{table}[h]
\caption{Evaluation of retrieval performance. P@k and R@k refer to Precision and Recall at k, respectively. Bolded values indicate the better-performing system for each metric. MITRA shows a clear advantage on the more realistic paraphrased queries (Set 2).}
\label{tab:rag-evaluation-1}
\centering
\begin{tabular}{llcccccc}
\toprule
\textbf{Query Set} & \textbf{System} & \textbf{P@1} & \textbf{R@1} & \textbf{P@3} & \textbf{R@3} & \textbf{P@5} & \textbf{R@5} \\
\midrule
\multirow{2}{*}{Set 1} & Keyword Search (BM25) & 1.00 & 0.85 & 0.40 & 0.90 & \textbf{0.32} & \textbf{1.00} \\
& MITRA & 1.00 & 0.85 & 0.40 & 0.90 & 0.24 & 0.90 \\
\midrule
\multirow{2}{*}{Set 2} & Keyword Search (BM25) & 0.13 & 0.03 & 0.25 & 0.56 & 0.18 & 0.59 \\
& MITRA & \textbf{0.75} & \textbf{0.66} & \textbf{0.33} & \textbf{0.81} & \textbf{0.20} & \textbf{0.81} \\
\bottomrule
\end{tabular}
\end{table}

While the P@k and R@k metrics in Table~\ref{tab:rag-evaluation-1} measure retrieval completeness, they do not fully capture the quality of the ranking. This is because a system that ranks the correct answer at position 1 receives the same P@5 score as a system that ranks it at position 5, they are rank-agnostic. Therefore, judging the system's quality on a P@k score alone can be misleading, as it does not reflect where the correct answers are placed in the list.

To provide a more comprehensive assessment of ranking quality---which is an important determinant of hallucination behavior in RAG---we evaluated both systems using rank-aware metrics: Mean Reciprocal Rank (MRR) and Normalized Discounted Cumulative Gain (NDCG), shown in Table \ref{tab:rag-evaluation-2}. On the keyword-based queries, both systems proved equally effective, achieving an identical MRR of 1.00. For the more challenging conceptual queries, MITRA's advantage is clear. It attains an MRR of 0.81 (compared to 0.35 for BM25) and an NDCG@5 of 0.88 (compared to 0.59 for BM25). This demonstrates two key advantages: the high MRR indicates the first correct answer is almost always ranked at the top, while the high NDCG confirms the overall list of retrieved passages is high-quality and correctly ordered. This combination ensures the generator conditions primarily on appropriate context, even when additional lower-ranked passages are retrieved.

\begin{table}[h]
\caption{Evaluation of retrieval performance. MRR reflects the average rank of the first correct answer, and NDCG@k measures the overall ranking quality of the top k retrieved answers. Bolded values indicate the better-performing system for each metric. MITRA shows a clear advantage on the more realistic paraphrased queries (Set 2).}
\label{tab:rag-evaluation-2}
\centering
\begin{tabular}{llcccccc}
\toprule
\textbf{Query Set} & \textbf{System} & \textbf{MRR} & \textbf{NDCG@3} & \textbf{NDCG@5} \\
\midrule
\multirow{2}{*}{Set 1} & Keyword Search (BM25) & 1.00 & 1.00 & 0.98  \\
& MITRA & 1.00 & 1.00 & \textbf{1.00}  \\
\midrule
\multirow{2}{*}{Set 2} & Keyword Search (BM25) & 0.35 & 0.67 & 0.59  \\
& MITRA & \textbf{0.81} & \textbf{0.91} & \textbf{0.88} \\
\bottomrule
\end{tabular}
\end{table}

Beyond these quantitative metrics, we also qualitatively tested the system's robustness against out-of-context queries. For example, when the system was locked on to a dark matter search analysis and asked, ``How many Higgs bosons were discovered in this search?", it did not hallucinate an answer. Instead, it correctly inferred from the retrieved passages that the document was unrelated to Higgs bosons and informed the user that the analysis in question is a dark matter search. This demonstrates the model's ability to effectively use the provided context to avoid generating nonsensical or misleading answers.

\section{Future Directions and Conclusion}
The current prototype demonstrates the feasibility and value of MITRA for knowledge retrieval from the physics collaboration database. While the evaluation in previous section has focused on rigorously benchmarking the retrieval component---the critical first step for mitigating hallucination---our immediate future work will focus on two key areas. First, we will expand the knowledge base to include multiple document types. Second, we will build a bigger and broader evaluation framework encompassing the generation step. This will include quantitative, LLM-as-a-judge metrics to systematically score faithfulness and answer relevancy, and will be supplemented by a formal user study with multiple domain experts to measure the system's performance and inter-annotator agreement.


Furthermore, we will conduct a formal performance benchmark to quantify query latency and throughput under realistic load. The current prototype exhibits low per-query latency (a few seconds as can be seen in the demo) and can process simultaneous user requests independently, suggesting that scaling to higher concurrency primarily requires increased compute capacity rather than architectural redesign. Future work includes deploying the system on a production-grade GPU cluster to handle the high-concurrency needs of a large collaboration and adding support for multi-turn conversational interactions to enable richer, context-aware assistance. Additionally, we will productionize the system by adopting a high-performance pluggable inference engine (e.g., vLLM or llama.cpp) and implementing strict network and access-control policies to ensure scalable, private deployment.

The ultimate vision is to evolve MITRA from a question-answering tool into a proactive research agent. Such an agent could not only retrieve information but also perform tasks like summarizing recent analysis updates, comparing methodologies between two related measurements, identifying potential strategies to handle disagreements between experimental data and Monte Carlo simulations, finding signal models with a similar topology, or finding gaps in current search spaces.

In conclusion, we have presented a novel framework for a RAG-based system tailored to the specific needs of large scientific collaborations. By combining an easily updatable, automated ingestion pipeline with a thoughtful, context-aware database design, and a commitment to data privacy, we provide a powerful tool to make research more efficient, accelerate the onboarding of new members, and unlock the full value of the collaboration's collective knowledge.


\textbf{Acknowledgement:} The authors wish to acknowledge that this work was supported by U.S. DOE, Office of Science, Office of High Energy Physics, under Award No. DE-SC0017647.

\printbibliography

\end{document}